\documentclass[twocolumn,twoside,preprintnumbers,amsmath,amssymb,showkeys]{revtex4}
\usepackage{epsfig}
\usepackage{graphicx}

\usepackage{fancyhdr}

\usepackage{pslatex}

\newcommand{\sqrtsNN}{\mbox{$\sqrt{\mathrm{s}_{_{\mathrm{NN}}}}$}}

\newcommand{\pt}{$p_{\rm T}$}
\newcommand{\DDbar}{${\rm D}\overline{\rm D}$}
\def \auau  {Au+Au}

\begin{document}

\title{\DDbar\ Correlations as a Sensitive Probe for Thermalization
in High-Energy Nuclear Collisions}

\author{
K.~Schweda$^{1}$,
X.~Zhu$^{2,3,4}$,
M.~Bleicher$^{2}$,
S.L.~Huang$^{5}$,
H.~St\"ocker$^{2,3}$,
N.~Xu$^{5}$,
P.~Zhuang$^{4}$
}
\affiliation{$^1$ Physikalisches Institut, Universit\"at Heidelberg, Philosophenweg 12, D-69120 Heidelberg, Germany}
\affiliation{$^2$ Institut f\"ur Theoretische Physik, Johann Wolfgang Goethe-Universit\"at, Max-von-Laue-Str.~1, D-60438 Frankfurt am Main, Germany} 
\affiliation{$^3$ Frankfurt Institute for Advanced Studies (FIAS), Max-von-Laue-Str.~1, D-60438 Frankfurt am Main, Germany}
\affiliation{$^4$ Physics Department, Tsinghua University, Beijing 100084, China}
\affiliation{$^5$ Nuclear Science Division, Lawrence Berkeley National Laboratory, Berkeley, CA 94720, USA}


\begin{abstract}
We propose to measure azimuthal correlations of heavy-flavor hadrons to address
the status of thermalization at the partonic stage of light quarks and
gluons in high-energy nuclear collisions.
In particular, we show that hadronic interactions at the late stage
cannot significantly disturb the initial back-to-back azimuthal correlations of \DDbar\
pairs. Thus, a decrease or the complete absence of these initial
correlations does indicate frequent interactions of heavy-flavor
quarks and also light partons in the partonic stage, which are essential
for the early thermalization of light partons.

\end{abstract}
\maketitle

\thispagestyle{fancy}

\setcounter{page}{1}

\section{Introduction}
Lattice QCD calculations, at vanishing or finite net-baryon density,
predict a cross-over transition from the deconfined thermalized partonic
matter (the Quark Gluon Plasma, QGP) to hadronic matter at a critical
temperature $T_{\rm c} \approx 150$--180~MeV~\cite{karsch}.

Measurements of hadron yields in the intermediate and high transverse
momentum (\pt) region indicate that dense matter is produced in
\auau\ collisions at RHIC~\cite{star_white,phenix_white}. The
experimentally observed large amount of elliptic flow of multi-strange
hadrons~\cite{star_omgeav2}, such as the $\phi$ meson and the $\Omega$
baryon, suggest that collective motion develops in
the early partonic stage of the matter produced in these collisions. A
crucial issue to be addressed next 
is the thermalization status of this partonic matter.

\begin{figure}[h]
\begin{center}
\includegraphics[width=0.3\textwidth]{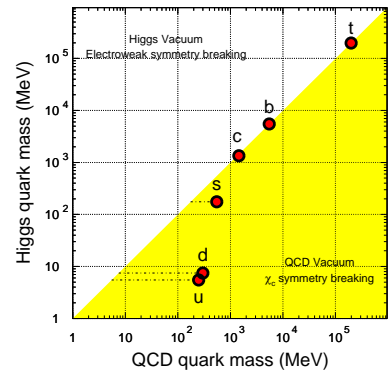}
\caption{(Color online) Quark masses in the QCD vacuum and the Higgs vacuum. A
  large fraction of the light quark masses is due to the chiral
  symmetry breaking in the QCD vacuum. The numerical values
  were taken from Ref.~\cite{pdg:2004}.} 
\label{fig1}
\end{center}
\end{figure}

Heavy-flavor (c, b) quarks are particularly excellent tools~\cite{Svetitsky:1996nj,
Svetitsky:1997xp,Svetitsky:1997bt} to study the
thermalization of the initially created matter. As shown in
Fig.~\ref{fig1}, their large masses are almost exclusively generated
through their coupling to the Higgs field in the electro-weak sector,
while masses of light quarks (u, d, s) are dominated by spontaneous
breaking of chiral symmetry in QCD. This means that in a QGP, where
chiral symmetry might be restored, light quarks are left with their bare
current masses while heavy-flavor quarks remain heavy. Due to their
large masses ($\gg\Lambda_{QCD}$), the heavy quarks can only be
pair-created in early stage pQCD processes. Furthermore, their production cross
sections in nuclear collisions are found to scale with the number of
binary nucleon-nucleon collisions~\cite{Adler:2004ta,starcharm_new}. In the subsequent
evolution of the medium, the number of heavy quarks is conserved because the
typical temperature of the medium is much smaller than the heavy
quark (c, b) masses, resulting in negligible secondary pair production. In addition,
the heavy quarks live much longer than the lifetime of the
formed high-density medium, decaying well outside. These
heavy quarks (c, b) can participate in collective motion or even
kinetically equilibrate if, and only if, interactions at the partonic level
occur at high frequency. The idea of statistical hadronization of
kinetically equilibrated charm quarks~\cite{pbm_3} even predicted
significant changes in hidden charm hadron
production~\cite{pbm_charm}.  
Hence, heavy-flavor quarks are an ideal probe to
study early dynamics in high-energy nuclear collisions.
Recent STAR and PHENIX results on elliptic flow and nuclear
modification factors of non-photonic electrons indicate that charm
quarks might indeed participate in the collective motion of the matter
produced in Au+Au collisions at RHIC.
To explain the data, a
large drag diffusion coefficient in the Langevin equation, describing
the propagation of charm quarks in the medium, is found to be
necessary \cite{Moore:2004tg}.  Two- and three-body
interactions~\cite{wicks_05,liu_06} of heavy-quarks and resonant
rescattering~\cite{hees_05} in the partonic stage seem to become important.
These investigations suggest that heavy-quarks actively participate in
the partonic stage.

In this paper, we study the change of azimuthal correlations of 
D and $\overline{\rm D}$ meson pairs as a sensitive indicator of
frequent occurrences of partonic scattering. Since heavy-flavor quarks
are pair-created by initial scattering processes, such as $gg\rightarrow
{\rm c}\overline{\rm c}$, each quark-antiquark pair is correlated in
relative azimuth $\Delta \phi$ due to momentum conservation. In
elementary collisions, these correlations survive the fragmentation
process to a large extent and hence are observable in the 
distribution of the relative azimuth of
pairs of D and $\overline{\rm D}$ mesons~\cite{hermine}.  We
evaluate by how much these correlations should be
affected by the early QGP stage and by the hadronic scattering processes
in the late hadronic stage.

\section{Results and Discussions}
To explore how the QCD medium generated in central ultra-relativistic
nucleus-nucleus collisions influences the correlations of D and
$\overline{\rm D}$ meson pairs, we employ a non-relativistic Langevin approach
which describes the random walk of charm quarks in a QGP as was first
described in
Refs.~\cite{Svetitsky:1996nj,Svetitsky:1997xp,Svetitsky:1997bt}.
Here, both the drag coefficient
$\gamma$ and the momentum-space diffusion coefficient $\alpha$ depend
on the \emph{local} temperature, $T$ with a momentum-independent 
$\gamma(T) = a T^2$. We chose $a = 2\cdot 10^{-6}$ (fm/$c$)$^{-1}$ MeV$^{-2}$.
Details of our calculations can be found in~\cite{ddbarCorr2006}. 

In order to isolate the effects purely due to parton-parton re-scattering in the
outlined medium, we generated c and $\overline{\rm c}$ quarks with the same $p_{\rm T}$ and zero longitudinal momentum back-to-back, i.e.\ with
$\Delta \phi = \pi$ at a time of the order of
$1/m_c\simeq$~0.1~fm/$c$, and used a delta function for fragmenting the
charm quark into a charmed hadron at the hadronization stage.  The
radius $r$ where each pair is created is randomly generated from a distribution
reflecting the number of binary nucleon-nucleon collisions taking
place at that radius, $p(r){\rm d}\,r\propto(R^2-r^2)\,2\pi\, r\, {\rm d}r$.
The angular distribution of the initially produced charm quarks
in the transverse plane is isotropic.

\begin{figure}[h]
\begin{center}
\includegraphics[width=0.4\textwidth]{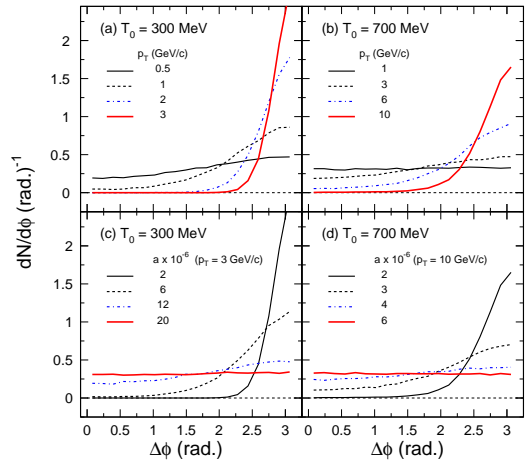}
\caption{(Color online) Correlations in relative azimuth $\Delta \phi$
  of \DDbar\ pairs from Langevin calculations with
  $T_0=300$~MeV (RHIC) and $700$~MeV (LHC).  The upper part shows the
  dependence of the correlations on the initial $p_{\rm T}$ of the c quark; 
  the lower part shows the drag
  coefficient dependence.}
\label{fig3}
\end{center}
\end{figure}

The evolution of the charm momenta with the Langevin equation is
stopped when $T$ reaches the critical temperature $T_{\rm c}=165$~MeV
or when the charm quark leaves the QGP volume. Furthermore, to get a first estimation
of the QGP effect on the charm quark pairs azimuthal correlation, we
omit the possible contribution of the mixed
phase. Figure~\ref{fig3}\,(a) shows the results for the D meson (charm
quark) pairs angular correlations for different initial charm quark $p_{\rm
T}$ values, for $T_0=300$~MeV and
$\tau_0=0.5$~fm/$c$ (typical values for RHIC collisions). 
We see that the fastest charm quarks (represented by the $p_{\rm
T}=3$~GeV/$c$) are able to escape from the QGP without suffering
significant medium effects, while the slower quarks (see the $p_{\rm
T}=0.5$~GeV/$c$ black line) have their pair azimuthal correlation almost
completely smeared out by the interactions in the medium.

Figure~\ref{fig3}\,(b) shows the corresponding results for $T_0=700$~MeV and
$\tau_0=0.2$~fm/$c$, values representative of LHC energies. Here, the interactions of
the charm quarks with the medium are so frequent that only the most
energetic charm quarks preserve part of their initial angular
correlation; low $p_{\rm T}$ pairs can even be completely stopped by
the medium. Because the c and the $\overline{\rm c}$ quarks of a given
pair are created together, in the
same space point, the pair has a higher escape probability
if both quarks escape the medium from the side of 
the fireball where the 
thickness is smaller. Thus, the $\Delta \phi$ distribution is shifted towards
smaller values. 


Presently, reliable numbers on the drag coefficient do not 
exist~\cite{Svetitsky:1987gq,hees_05}.
The dependence on the drag coefficient is shown in
Figs.~\ref{fig3}\,(c) and (d) for the c$\overline{\rm c}$ angular correlations
for energetic c quarks: $p_{\rm T}=3$ and 10~GeV/$c$ for $T_0=300$ and
700~MeV, respectively.  These correlations vanish when $a$ is increased 
by around a factor of five in the first
case and by around a factor of two in the latter one, with respect to values
from pQCD-calculations.

We have demonstrated that initial correlations of ${\rm c}\overline{\rm c}$ pair correlations are
clearly affected by the formed QCD medium, and sensitive to
its temperature and drag coefficient. 
The question is to what extent hadronic interaction can modify
these initial correlations. 
In the following, we apply the microscopic hadronic
transport approach UrQMD v.~2.2~\cite{urqmd1,urqmd2}.
A phase transition to a QGP state is \emph{not} incorporated
into the model dynamics. In the latest version of the model,
PYTHIA (v.~6.139) was integrated to describe the energetic primary
elementary collisions~\cite{Bratkovskaya:2004kv}. Measured particle
yields and spectra  
are well reproduced by this approach~\cite{Bratkovskaya:2004kv}.

\begin{figure}[h]
\vspace{-0.5cm}
\begin{center}
\includegraphics[width=0.45\textwidth]{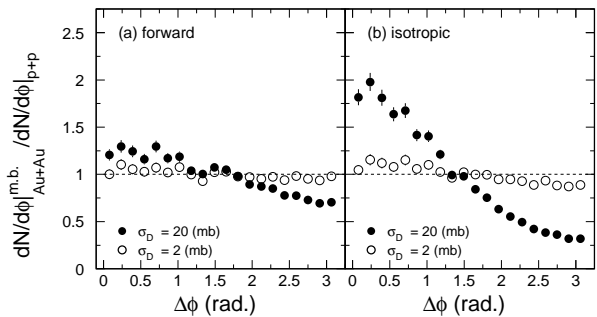}
\vspace{-0.6cm}
\caption{Ratios between the $\Delta \phi$ distributions of \DDbar\
  pairs produced in minimum bias Au+Au reactions at \sqrtsNN$=200$~GeV, before and
  after hadronic rescattering, using forward (left panel) and
  isotropic (right panel) angular distributions.} 
\label{fig4}
\end{center}
\end{figure}

In this model, D mesons stem from the early-stage high-energy
nucleon-nucleon collisions, calculated with PYTHIA. For their
propagation in the hadronic medium, we consider elastic scattering of
D mesons with all other hadrons. The hadronic scattering cross-section
for D mesons is generally considered to be small~\cite{Lin:2000jp} and we take 2~mb in
our calculation. 
The results of the UrQMD calculations are shown in Fig.~\ref{fig4}\,(a),
for minimum bias Au+Au reactions at \sqrtsNN$=200$~GeV. To better illustrate the
\emph{change} in the angular correlations of the ${\rm D}\overline{\rm
D}$ pairs, we show the ratio between the final distribution, affected
by the evolution with UrQMD, and the initial one, directly obtained
from PYTHIA. In case hadronic interactions do not modify the 
correlations, this ratio is unity. 

For this specific analysis, all
${\rm D}\overline{\rm D}$ pairs were selected, irrespective of the
$p_{\rm T}$ of the D mesons. Fig.~\ref{fig4}\,(a) shows our results assuming a forward peaked
scattering cross section, $\propto \exp(7\cos\theta)$, 
where $\theta$ is the scattering angle in the center of mass system of the
decaying resonance. Even for a large cross section of 20~mb, the ${\rm D}\overline{\rm
D}$ pair correlations are barely affected by hadronic interactions. 

On the other hand, intermediate resonances formation (essentially
D$+\pi\leftrightarrow$D$^*$), might result in isotropic emission, 
i.e. the correlation might be completely destroyed by a single scattering
process. Only when assuming the unlikely case of a large cross section    
of 20~mb with fully isotropic emission, the ${\rm D}\overline{\rm D}$ pair 
correlations are significantly modified, see Fig.~\ref{fig4}\,(b).
We also find that most scatterings occur at relatively early times, when
the hadron density is high. At RHIC energies, the (pure) hadronic
stage is presumably shorter than assumed in the present calculations.

Hence, we conclude that hadronic interactions are unlikely to significantly 
modify ${\rm D}\overline{\rm D}$ pair correlations. 
A change in these angular correlation must be dominated by frequent parton-parton
scatterings occurring in the QGP phase.

The calculations reported above show that heavy-flavor correlations
provide a sensitive tool to directly access the status of
equilibration of the partonic medium. A similar picture emerges from the
observation and suppression of back-to-back correlations of
unidentified high $p_{\rm T}$ hadrons (jets).  Both features originate
from the propagation of partons (heavy quarks or jets) in the medium.
However, D mesons (and B mesons) have the advantage that they can
still be identified, even if they lose a significant fraction of their
energy and get kinetically equilibrated. 

\section{Conclusion and Outlook}
In summary, we argue that the observation of broadened angular
correlations of heavy-flavor hadron pairs in high-energy
heavy-ion collisions would be an indication of thermalization at the
partonic stage (among light quarks and gluons). We have seen that
hadronic interactions at a late stage in the collision evolution
cannot significantly disturb the azimuthal correlations of \DDbar\
pairs.  Thus, a visible decrease or the complete absence of such
correlations, would indicate frequent interactions of heavy-flavor
quarks and other light partons in the partonic stage, implying early thermalization
of light quarks in nucleus-nucleus collisions at RHIC and LHC.

These measurements require good statistics of events in which
\emph{both} D mesons are cleanly reconstructed. A complete
reconstruction of the D mesons (i.e.\ of \emph{all} their decay
products) in full azimuth is essential to preserve the kinematic
information and to optimize the acceptance for detecting correlated D
meson pairs. Solid experimental measurements in $pp$ and light-ion
collisions, at the same energy, are crucial for detailed studies of
changes in these azimuthal correlations, and should be performed as a
function of \pt. Proposed upgrades of STAR~\cite{star_hft} 
and PHENIX at RHIC and the 
ALICE detector~\cite{alice_ppr} at LHC with micro-vertex capabilities 
and direct open charm reconstruction should make 
these measurements possible.

\noindent{\bf Acknowledgements} 
K.S. acknowledges support by the Helmholtz foundation 
under contract number VH-NG-147. This work has been supported by GSI and BMBF, 
in part by the U.S. Department of Energy under Contract No. DE-AC03-76SF00098.


\end{document}